\documentclass[11pt]{article}
\usepackage{latexsym,amssymb}
\newcommand{\ft}[2]{{\textstyle\frac{#1}{#2}}}
\def\tilde{\widetilde}

\def\1bar{1\hskip -.275cm -}
\def\2bar{2\hskip -.275cm -}
\def\3bar{3\hskip -.275cm -}

\newsavebox{\uuunit}
\sbox{\uuunit}
    {\setlength{\unitlength}{0.825em}
     \begin{picture}(0.6,0.7)
        \thinlines
        \put(0,0){\line(1,0){0.5}}
        \put(0.15,0){\line(0,1){0.7}}
        \put(0.35,0){\line(0,1){0.8}}
       \multiput(0.3,0.8)(-0.04,-0.02){10}{\rule{0.5pt}{0.5pt}}
     \end {picture}}

\makeatletter \@addtoreset{equation}{section} \makeatother

\def\bfone{\relax{\rm 1\kern-.35em 1}}

\def\bfone{\relax{\rm 1\kern-.35em 1}}
 
\begin{document}
\begin{titlepage}
\begin{flushright}
DFTT - 41/2002\\
\end{flushright}
\vskip 1.5cm
\begin{center}
{\LARGE \bf  An  $\mathrm{SL(2,\mathbb{R})}$-covariant, first order,
 $\kappa$-supersymmetric  action for
  the $D5$-brane $^ \dagger $}\\ \vfill {\large
 Pietro Fr\'e and  Leonardo Modesto} \\
\vfill {
 Dipartimento di Fisica Teorica, Universit\'a di Torino, \\
$\&$ INFN -
Sezione di Torino\\
via P. Giuria 1, I-10125 Torino, Italy  }
\end{center}
\vfill
\begin{abstract}
{The new first order, rheonomic, $\kappa$--supersymmetric
formalism recently introduced by us for the  world-volume action of
the $D3$ brane is extended to the case of $D5$ branes.  This
extension requires the dual formulation of the Free Differential
Algebra of type IIB supergravity in terms of $6$--form gauge
potentials which was so far missing and is given here. Furthermore
relying on our new approach we are able to write the $D5$ world
volume action in a manifestly $\mathrm{SL(2,\mathbb{R})}$ covariant
form. This is important in order to solve the outstanding problem of
finding the appropriate boundary actions of $D3$--branes on smooth
ALE manifolds with twisted fields. The application of our results to
this problem is however postponed to a subsequent publication.}
\end{abstract}
\vspace{2mm} \vfill \hrule width 3.cm {\footnotesize $^ \dagger $
This work is supported in part by the European Union RTN contracts
HPRN-CT-2000-00122 and HPRN-CT-2000-00131.}
\end{titlepage}
\section{Introduction}
In a recent paper \cite{noidued3}, the present authors introduced
a \textbf{new first order formalism} that is able of generating second order world volume actions of
the Born--Infeld type for $p$-branes.
This construction allows for a description within the
standard framework of rheonomy of those $\kappa$-supersymmetric boundary actions
that function as sources for  supergravity $p$--solutions in the bulk. Furthermore
in this approach the $\mathrm{SL(2,\mathrm{R})}$
duality symmetry  can be manifest at all levels.
These features are very much valuable and quite essential in order
to discuss various questions relative to the microscopic
interpretation of classical $p$--solutions like the $D3$--brane with flux on
smooth ALE manifolds \cite{noialtrilast} and other issues in the
general quest of the gauge/gravity correspondence.
The new formulation which, in our opinion, is
particularly compact and elegant is based on the introduction of an
additional auxiliary field, besides the world volume vielbein, and on
the enlargement of the local symmetry from the Lorentz group to the
general linear group:
\begin{equation}
  \mathrm{SO(1,d-1)}\,  \stackrel{\mbox{enlarged}}{\Longrightarrow}
  \, \mathrm{GL(d,\mathbb{R})}
\label{allargo}
\end{equation}
Within this new formalism in \cite{noidued3} we constructed the first order
version of the $\kappa$--supersymmetric action for
a $D3$--brane. This choice was not random rather it was rooted in the main
motivations to undertake such a  new construction. Indeed we needed a
suitable formulation of the $D3$--brane action holding true on a
generic background in order to apply it to the aforementioned
$D3$--brane bulk solution with flux \cite{noialtrilast} or to more complicated ones.
However, as for the the solution in \cite{noialtrilast}, the task was not
exhausted once we had the $D3$-brane action. Indeed this latter accounts
only  for the source of the  Ramond-Ramond $A^{[4]}$--form with self dual field
strength but not for the source of the doublet $\{ B^{[2]} , C^{[2]} \}$
of $2$--forms whose flux is trapped on the homology $2$--cycles of
the transverse ALE manifold. The source of such fields can only be accounted for
by the presence of $5$--branes whose world-volume sweeps both the
world volume of the $D3$--brane and the ALE $2$--cycles. Hence we
need to extend our first order formalism also to $5$--branes and this
is what we do in the present paper.
\par
It was noted already in \cite{copetorino1,noialtrilast,copetorino2}
that the a $D5$ brane wrapped on a $2$--cycle of the ALE manifold
cannot be the direct source of the smooth, fractional brane like
solution of \cite{noialtrilast}. The reason is simple. In the
solution of \cite{noialtrilast} the dilaton field is constant and
therefore there is no source for its field equation. On the other
hand the $D5$--brane couples to the dilaton and contributes such a
source. It is only in the orbifold limit that such a coupling
vanishes together with the vanishing of the homology cycle trapping
the flux of the $2$--forms. In order to reproduce the correct
boundary action for the smooth supersymmetric boundary solution of
\cite{noialtrilast} we need a $5$--brane that has vanishing coupling
to the dilaton field. As we are going to show in a next coming
publication \cite{noiprossimo} such a brane exists and it is a mixture of $D5$--branes
on which closed \textbf{fundamental strings} can annihilate and of
$D5^\prime$--branes where such annihilation occurs for closed
\textbf{$D$-strings}. To describe the world--volume action of such an
object we need an explicitly $\mathrm{SL(2,\mathbb{R})}$-covariant
formulation of the $5$ brane that depends on two vectors $q_{\Lambda}$ and $p_{\Lambda}$ that transform in the fundamental representation of $\mbox{SL}(2, \mathbb{R})$.
In this paper using the first order rheonomic formalism of
\cite{noidued3} and relying on a dual formulation of type IIB
supergravity we construct the required $\mathrm{SL(2,\mathbb{R})}$
covariant $5$--brane action.
\section{Type IIB supergravity with dual $6$--form potentials}
In the appendix $B$ of \cite{noidued3} we gave
a short but comprehensive summary of type IIB supergravity in the
rheonomic approach. This summary was entirely based on the original papers by
Castellani and Pesando \cite{castella2b,igorleo} but it also contained
some new useful results, in particular the transcription of the
curvatures from the complex $\mathrm{SU(1,1)}$ basis to the real
$ \mathrm{SL(2,\mathbb{R})}$ basis and a comparison of the
supergravity field equations as written in the rheonomy approach and
as written in string text--books like Polchinsky's \cite{polchinski}.
\par
In order to deal with $5$--branes we need to include also the
$7$--form field strengths that are dual to the $3$--form field
strengths originally used in \cite{castella2b,igorleo} and summarized
in \cite{noidued3}. This amounts to extending the original Free
Differential Algebra of type IIB supergravity with the $6$--form
generators that are associated with the new cohomology classes of
such an algebra. This is the exact analogue of the procedure which
leads to the extended  Free Differential Algebra of M-theory
\cite{pietd11} containing both
the $3$--form (source of $M2$-branes) and the $6$--form (source of the $M5$-branes). This latter arises
by including the new generator associated with the cohomology $7$--cocycle of the
original Free Differential Algebra of M--theory in its $3$--form
formulation \cite{riccapiet}.
\par
In our case the new complete form of the curvatures is the following
one:
\begin{eqnarray}
   R^{\underline{a}}&=&{\cal D} V^{\underline{a}}-i{\bar\psi}\wedge \Gamma^{\underline{a}}\psi \label{tors11}\\
   R^{\underline{ab}}&=&d \omega^{\underline{ab}}-\omega^{\underline{ac}} \wedge\omega^{\underline{db}} \,
   \eta_{\underline{cd}}\cr
   \rho&=&{\cal D} \psi \equiv d\psi- \ft 1 4  \omega^{\underline{ab}} \wedge \Gamma_{ab}\psi-
          \ft 1 2 {\rm i} Q \wedge \psi \label{lore11}\\
   \mathcal{H}^{\alpha}_{[3]} &=& \sqrt{2} \, dA^{\alpha}_{[2]}+2i\Lambda^{\alpha}_+ {\bar\psi}
   \wedge \Gamma_{\underline{a}} \psi^* \wedge V^{\underline{a}}
          +2{\rm i}\Lambda^{\alpha}_- {{\bar\psi}}^* \wedge
          \Gamma_{\underline{a}} \psi \wedge V^{\underline{a}}\label{2form11}\\
   \mathcal{F}_{[5]} &=& dC_{[4]}+\ft {1} {16} \, {\rm i} \,
   \epsilon_{\alpha\beta} \, \sqrt{2} \, A^{\alpha}_{[2]}  \wedge \mathcal{H}^\beta_{[3]} \,+\, \ft 1 6 \,
   {\bar\psi} \wedge
   \Gamma_{\underline{abc}}\psi \wedge V^{\underline{a}}  \wedge V^{\underline{b}} \wedge
   V^{\underline{c}}
    \nonumber\\
&& + \, \ft 1 8 \, \epsilon_{\alpha\beta} \, \sqrt{2} \, A^{\alpha}_{[2]}
\wedge\left( \Lambda^\beta_+ {\bar\psi} \Gamma_{\underline{a}} \psi^\star +
 \Lambda^\beta_- {\bar\psi}^\star \Gamma_{\underline{a}}
\psi\right) \,
 \wedge V^{\underline{a}} \label{4form11}\\
   {\cal D}\lambda &=& d\lambda-{1\over 4} \omega^{\underline{ab}} \Gamma_{\underline{ab}}\lambda
    -{\rm i}  \ft 3 2 \, Q\lambda \label{dilatin11} \\
  {\cal D}\Lambda^{\alpha}_\pm&=&d\Lambda^{\alpha}_{\pm} \mp
   {\rm i}  \, Q \, \Lambda^{\alpha}_{\pm}.\label{su11mc11} \\
\mathcal{H}^{\alpha}_{[7]} & = & dC_{[6]}^{\alpha} + \Lambda^{\alpha}_{+} \,
{\bar{\psi}} \, \Gamma_{a_1 a_2 a_3 a_4 a_5} \, \psi^{*} \,
V^{a_1 a_2 a_3 a_4 a_5} - \Lambda^{\alpha}_{-} \,
{\bar{\psi}}^{*} \, \Gamma_{a_1 a_2 a_3 a_4 a_5} \, \psi \,
V^{a_1 a_2 a_3 a_4 a_5}  \nonumber \\
&& + \frac{10}{3} \, {\bar{\psi}} \,
\Gamma_{a_1 a_2 a_3} \, \psi \, V^{a_1 a_2 a_3} \, \sqrt{2} \, A^{\alpha}_{[2]} \nonumber \\
&& + 40 \, \sqrt{2} \, dA^{\alpha}_{[2]} \, C_{[4]} \label{h7curva}
\end{eqnarray}
the novelty being the expression of the $\mathcal{H}^{\alpha}_{[7]}$
curvature. Its definition is a direct consequence of the following
Fierz identities:
\begin{eqnarray}
\bar{\psi} \, \Gamma^{a} \, \psi^{*} \, \bar{\psi} \, \Gamma^{a} & = & 0 \nonumber \\
\Gamma^{a} \, \psi \, \bar{\psi}^{*} \, \Gamma^{a} \, \psi & = & 0 \nonumber \\
\bar{\psi} \, \Gamma_{a_1 a_2 a_3} \, \psi \, \bar{\psi} \, \Gamma^{a_3} & = & \bar{\psi}^{*} \, \Gamma_{[a_1} \, \psi \, \bar{\psi}_{a_2]} \, \psi^{*} \nonumber \\
\bar{\psi} \, \Gamma_{a_1 a_2 a_3 a_4 a_5} \, \psi^{*} \, \bar{\psi} \, \Gamma^{a_1} & = & - \, 4 \, \bar{\psi} \, \Gamma_{[a_2 a_3 a_4} \, \psi \, \bar{\psi} \, \gamma_{a_5]} \, \psi^{*} \nonumber \\
\bar{\psi}^{*} \, \Gamma_{a_1 a_2 a_3 a_4 a_5} \, \psi \, \bar{\psi} \, \Gamma^{a_1} & = &  4 \, \bar{\psi} \, \Gamma_{[a_2 a_3 a_4} \, \psi \, \bar{\psi}^{*} \, \gamma_{a_5]} \, \psi
\end{eqnarray}
From eq.s (\ref{h7curva}) and (\ref{4form11}) one obtains the
supersymmetry transformation rules of the background fields relevant
in establishing the $\kappa$-supersymmetry transformation rules
against which the $5$-brane action is invariant.
\section{Establishing the $\kappa$-supersymmetric action of the $5$--brane}
\label{thed3examp}
Next, following the general scheme outlined in \cite{noidued3} we apply
the new first order formalism  to the case $d=6$ in order to derive the $\kappa$-supersymmetric
action of a $5$--brane. As extensively discussed in \cite{noidued3},
the $\kappa$--supersymmetry transformations just follow, via a suitable projection, from
the bulk supersymmetries as derived from supergravity, type IIB
theory, in this case. The latter has a duality symmetry with respect
to an $\mathrm{SL(2,\mathbb{R})}$ group of transformations that acts
non linearly on the two scalars of massless spectrum, the dilaton $\phi$ and
the Ramond scalar $C_0$. Indeed these two parametrize the coset
manifold $\mathrm{SL(2,\mathbb{R})}/\mathrm{O(2)}$ and actually
correspond to its solvable parametrization. Because of the special
relevance of this issue in the context of our problem, we recall below
such a solvable parametrization. This also helps to fix the notations.
\par
\leftline{\underline{\sl $\mathrm{SL(2,\mathbb{R})}$ Lie algebra}}
\begin{equation}
  \left[ L_0 \, , \, L_\pm \right] = \pm \, L_\pm \quad ; \quad \left[ L_+ \, , \, L_- \right] = 2 \,
  L_0
\label{sl2alg}
\end{equation}
with explicit $2$--dimensional representation:
\begin{equation}
  L_0= \ft 1 2 \, \left( \begin{array}{cc}
    1 & 0 \\
    0 & -1 \
  \end{array}\right) \quad ; \quad L_+= \left( \begin{array}{cc}
    0 & 1 \\
    0 & 0 \
  \end{array}\right) \quad ; \quad L_-= \left( \begin{array}{cc}
    0 & 0 \\
    1 & 0 \
  \end{array}\right) \quad ; \quad
\label{2drepsl2r}
\end{equation}
\par
\leftline{\underline{\sl Coset representative of $\mathrm{SL(2,\mathbb{R})/O(2)}$ in the solvable parametrization}}
\begin{equation}
  \mathbb{L}\left( \varphi , C_{[0]} \right) =\exp \left[  \varphi \, L_0 \right]  \, \exp \left[
  C_{[0]} e^{\varphi} \, L_-\right] \, = \, \left( \begin{array}{cc}
    \exp[\varphi/2] & 0 \\
    C_{[0]}e^{\varphi/2} & \exp[-\varphi/2] \
  \end{array}\right)
\label{Lfatcoset}
\end{equation}
where $\varphi(x)$ and $ C_{[0]}$ are respectively identified with
the dilaton and with the Ramond-Ramond 0-form of the superstring
massless spectrum.
The isomorphism of $\mathrm{SL(2,\mathbb{R})}$ with
$\mathrm{SU(1,1)}$ is realized by conjugation with the Cayley matrix:
\begin{equation}
  \mathcal{C}= \ft {1}{\sqrt{2}} \, \left( \begin{array}{cc}
    1 & - {\rm i} \\
    1 & {\rm i} \
  \end{array}\right)
\label{caylmat}
\end{equation}
Introducing the $\mathrm{SU(1,1)}$ coset representative
\begin{equation}
\mathrm{SU(1,1)}\,  \ni \,\Lambda \, = \, \mathcal{C} \, \mathbb{L}
\, \mathcal{C}^{-1}
\label{Lambigcos}
\end{equation}
from the left invariant $1$--form $\Lambda^{-1} \, d \Lambda $ we can
extract the $1$-forms corresponding to the scalar vielbein $P$ and the $\mathrm{U(1)}$ connection $Q$
\par
\leftline{\underline{\sl The $\mathrm{SU(1,1)/U(1)}$ vielbein and
connection}}
\begin{equation}
  \Lambda^{-1} \, d \Lambda \, = \, \left( \begin{array}{cc}
   - {\rm i}\, Q & P \\
    P^\star & {\rm i}\, Q \
  \end{array} \right)
\label{su11viel}
\end{equation}
Explicitly
\begin{equation}
\begin{array}{rcll}
P & = & \ft 1 2 \, \left( d\varphi - {\rm i}\, e^\varphi\, dC_{[0]} \right)  & \mbox{scalar vielbein} \\
Q & = & \ft 1 2 \,  \exp[ \varphi ] \, dC_{[0]} & \mbox{$\mathrm{U(1)}$-connection} \
\end{array}
\label{PQvalue}
\end{equation}
As stressed in the introduction the $5$--brane action we want to write, not only should be cast into
first order formalism, but should also display manifest covariance
with respect to $\mathrm{SL(2,\mathbb{R})}$. This covariance relies
on introducing two vectors with two components $q_\alpha$ and $p_\alpha$  that transform in the
fundamental representation of $\mathrm{SU(1,1)}$
and expresse the charges carried by the $D5$ brane with respect to
the $2$--forms $A^\alpha_{[2]}$ of bulk supergravity (both the Neveu Schwarz $B_{[2]}$ and
Ramond--Ramond $C_{[2]}$). According to the geometrical formulation of type IIB
supergravity summarized above we set:
\begin{equation}
\begin{array}{rclcrcl}
A^{\Lambda} & = & (B_{[2]} \, , \, C_{[2]}) &;&
A^{\alpha} & =& \mathcal{C}^{\alpha} \, \, _{\Lambda} \, A^{\Lambda}
\\
A^{\alpha  = 1} & = & \frac{1}{\sqrt{2}} \, (B_{[2]} \, - \, i \, C_{[2]})
&;&
A^{\alpha  = 2} & = & \frac{1}{\sqrt{2}} \, (B_{[2]} \, + \, i \, C_{[2]})\
\end{array}
\label{def A}
\end{equation}
In terms of these objects we write down the complete action of the $D5$--brane as follows:
\begin{eqnarray}
\mathcal{L} & = & \rho \, \Pi^{\underline{a}} _i \, V^{\underline{b}} \, \eta_{\underline{ab}}
 \, \eta^{i\ell_1} \, \wedge e^{\ell_2} \, \wedge \, \dots \wedge e^{\ell_6} \, \epsilon_{\ell_1 \dots \ell_6}
  + \rho \, a_1 \, \Pi^{\underline{a}} _i \, \Pi^{\underline{b}}
_j \, \eta_{\underline{ab}} \, h^{ij} \, e^{\ell_1} \, \wedge \, \dots
\wedge e^{\ell_6} \, \epsilon_{\ell_1 \dots \ell_6} \nonumber\\
&&+ \rho \, a_2 \, \left [ \mbox{det}\,\left(  h^{-1} + \mu \mathcal{F}\right) \right] ^\alpha \,
  e^{\ell_1} \, \wedge \, \dots
\wedge e^{\ell_6} \, \epsilon_{\ell_1 \dots \ell_6}\nonumber\\
&& + \rho \, a_3 \mathcal{F}^{ij} \, F^{[2]} \, \wedge \, e^{\ell_3} \, \wedge \,e^{\ell_6} \,
\epsilon_{ij\ell_3 \dots \ell_6} \nonumber \\
&&+\nu \, F \, \wedge \, F \, \wedge \, F + a_5 \, p_{ \alpha} \, A^{\alpha} \,
\wedge \, F \wedge \, F + a_6 \, C_{[4]} \wedge \, F + a_7 \, \tilde{q}_{\alpha} C_{[6]}^{\alpha}
\label{us eq}
\end{eqnarray}
where :
\begin{equation}
\quad q_\alpha \, p_\beta \, \epsilon^{\alpha\beta} = 1
\label{qvecti}
\end{equation}
$q_{\alpha} = \frac{1}{\sqrt{2}} (q_1 , q_1^{\star})$, $p_{\alpha} = \frac{1}{\sqrt{2}} (p_1 , p_1^{\star})$ (see appendix \ref{covariance}) 
and where $\rho \, = \, \rho(\phi)$ is a function of the dilaton to be determined,
$C_{[4]}$ is the $4$--form potential, and $C_{[6]}^{\alpha}$ are the dual potentials to
the $A_{[2]}^{\alpha}$ forms. The coefficients
\begin{equation}
\alpha\, = \, \frac{1}{4} \hspace{1cm} a_1 \, = \, - \, \frac{1}{12}  \hspace{1cm} a_2 \, = \, - \,
\frac{1}{3} \hspace{1cm}
  a_3\, = \, \frac{5}{4}
\label{repetita}
\end{equation}
where already determined in paper \cite{noidued3}. The coefficients $a_5,a_6,\nu$ are new
and they must be fixed by to be $\kappa$--supersymmetry. The first two
are numerical, while $\nu$ is  also a function  of the bulk scalars to be determined through
$\kappa$-supersymmetry.
\par
In the action (\ref{us eq})
\begin{equation}
  F^{[2]} \equiv F_{[G]} + q_{\alpha} A^{\alpha}
\label{nFdefidA}
\end{equation}
is the field strength of the world--volume gauge field and depends on
the charge vector $q_{\alpha}$. The physical interpretation of
$F^{[2]}$ is as follows. By definition a $Dp$--brane is a locus in
space--time where open strings can end or, in the dual picture,
boundaries for closed string world--volumes can be located. The type IIB theory contains
two kind of strings, the fundamental strings and the $D$-strings
which are rotated one into the other by the $\mathrm{SL(2,\mathbb{Z})}\subset
\mathrm{SL(2,\mathbb{R})} $ group. Correspondingly a $D5$ brane can be
a boundary either for fundamental or for $D$--strings or for a
mixture of the two. The charge vectors $q_\alpha$ and $p_{\alpha}$ just express this
fact. 
Furthermore the definition (\ref{nFdefidA}) of $F^{[2]}$
encodes the following idea: the world--volume gauge $1$-form $A^{[1]}$
is just the parameter of a gauge transformation for the $2$--form $q_{\alpha}
A^{\alpha}$, which in a space--time with boundaries can be reabsorbed
everywhere except on the boundary itself. Note that if we
take $q_{\alpha}  = \frac{1}{\sqrt{2}}(1 \, , \, 1)$ and $p_{\alpha}  = \frac{i}{\sqrt{2}}(1 \, , \,- 1)$ we obtain :
\begin{equation}
q_{\alpha} A^{\alpha} \, = \, B_{[2]} \quad ; \quad
   p_{\beta}  \, A^{\beta} \, = \, C_{[2]}
\label{no q}
\end{equation}

\subsection{$\kappa$--supersymmetry}
Next we want to prove that with an appropriate choice of $\nu, a_5$, $a_6$ and $a_7$ the action (\ref{us eq}) is invariant against bulk
supersymmetries characterized by a projected spinor parameter. For
simplicity we do this in the case of the choice $\tilde{q}_{\alpha} = q_{\alpha}  = \frac{1}{\sqrt{2}}(1 \, , \,1)$ and $p_{\alpha}  = \frac{i}{\sqrt{2}}(1 \, , \,- 1)$ . For other choices of the charge type the modifications needed in
the prove will be obvious from its details.
\par
To accomplish our goal we begin by writing the supersymmetry transformations of the
bulk differential forms $V^{\underline{a}}$, $B_{[2]}$, $C_{[2]}$,
$C_{[4]}$ and $C_{[6]}$ which appear in the action. From the rheonomic
parametrizations of the curvatures recalled in \cite{noidued3} we immediately
obtain:
\begin{eqnarray}
&&\delta V^{\underline{a}}\, = \, \mbox{i}\, \ft 1 2 \, \left( \overline{\epsilon} \,
\Gamma^{\underline{a}} \, \psi + \overline{\epsilon}^{*} \,
\Gamma^{\underline{a}} \, \psi^{*}\right)   \nonumber\\
&&\delta B_{[2]} \, = \, - \, 2 \, i \, [(\Lambda^{1}_{+} +\Lambda^{2}_{+}) \,
{\bar\epsilon} \, \Gamma_{\underline{a}} \, \psi^{*} \,
V^{\underline{a}}+(\Lambda^{1}_{-} +\Lambda^{2}_{-}) \, {\bar\epsilon}^{*} \,
\Gamma_{\underline{a}} \, \psi \, V^{\underline{a}}] \nonumber\\
&&\delta C_{[2]} \, = \, 2 \, [(\Lambda^{1}_{+} -\Lambda^{2}_{+}) \, {\bar\epsilon} \,
\Gamma_{\underline{a}} \, \psi^{*} \,
V^{\underline{a}}+(\Lambda^{1}_{-} -\Lambda^{2}_{-}) \, {\bar\epsilon}^{*} \,
\Gamma_{\underline{a}} \, \psi \, V^{\underline{a}}] \nonumber\\
&&\delta C_{[4]} \, = \underbrace{ -\frac{1}{6} ( {\bar\epsilon} \, \Gamma_{\underline{abc}}
\, \psi \, - \, {\bar\epsilon}^{*} \, \Gamma_{\underline{abc}} \, \psi^{*})
\, V^{\underline{abc}} }_{\delta C_{[4]}^{\prime}} + \frac{1}{8} \,  [ B_{[2]} \,
\delta C_{[2]} \, - \, C_{[2]} \, \delta B_{[2]} ]
\label{susy tran}
\end{eqnarray}
and :
\begin{eqnarray}
\delta C_{[6]} \, && = \, g_{1} \, C_{[4]} \, \delta F_{[2]} + \, g_{2} \,F_{[2]} \delta C_{[4]}^{\prime} + \delta C_{[6]}^{\prime} \nonumber \\
&& = \frac{1}{\sqrt{2}}(\delta C_{[6]}^{1} + \delta C_{[6]}^{2}) \nonumber \\
&& = - \frac{2}{\sqrt{2}}(\Lambda^{1}_{+} + \Lambda^{2}_{+}) {\bar\epsilon} \, \Gamma_{\underline{a_{1} \dots a_{5}}}\, \psi^{*} \, + \, \frac{2}{\sqrt{2}} (\Lambda^{1}_{-} + \Lambda^{2}_{-}) {\bar\epsilon}^{*} \, \Gamma_{\underline{a_{1} \dots a_{5}}} \, \psi \, V^{\underline{a_{1} \dots a_{5}}}+ \nonumber \\
&& \hspace{0.45cm} + 20 \, \sqrt{2} \, \delta C_{[4]}^{\prime} \, F_{[2]} \, - \, 40 \, \sqrt{2} \, C_{[4]} \, \delta F_{[2]}
\label{var.ci.6}
\end{eqnarray}
Note that in writing the above transformations we have neglected all
terms involving the dilatino field. This is appropriate since the
background value of all fermion fields is zero. The gravitino
$1$--form $\psi$ is instead what we need to keep track of. Proving
$\kappa$--supersymmetry is identical with showing that all $\psi$
terms cancel against each other in the variation of the action.
Relying on (\ref{susy tran}) the variation of the W.Z.T term is as follows:
\begin{eqnarray}
&&\delta ( \nu \, F \wedge \, F \, \wedge F +a_5 \, C_{[2]} \wedge F \, \wedge \, F + a_{6} \, C_{[4]} \, \wedge F + a_{7} \, C_{[6]}) \, = \nonumber\\
&&= 3 \, \nu \, F \, F \, \delta F - \frac{3 g_{1} a_{7}}{16}  \, F \, F \, \delta C_{[2]} + (g_{2} - g_{1}) a_{7} \, F \, \delta C_{[4]}^{\prime} + a_{7} \, \delta C_{[6]}^{\prime}
\label{var wzt}
\end{eqnarray}
for $a_5 \, = \, - \frac{g_1 a_7}{16}$ , $a_6 \, = \, - g_1 a_7$ , $g_1 \, = \, - \, 40 \, \sqrt{2}$ and $g_2 \, = \, 20 \, \sqrt{2}$ .\\
And with such a choice the complete variation of the Lagrangian under a supersymmetry  transformation of arbitrary parameter is:
\begin{eqnarray}
\delta \mathcal{L} \, & = & \, \delta \mathcal{L}_{\psi} +  \delta \mathcal{L}_{\psi^{*}} \nonumber\\
\delta \mathcal{L}_{\psi} \, & = & \, [- \, 5! \, i \, (h^{-1})^{ij} \, ({\bar\epsilon} \, \gamma_{i} \, \psi ) \, + \,
(\mu_{3} \, \mathcal{F}^{ij} \, + \, \mu_{4} \, {\tilde{F \, F}}^{ij}) \, ({\bar\epsilon}^{*} \, \gamma_{i} \, \psi ) \, + \, \nonumber\\
&& + (g_1 - g_2) \frac{a_7}{3} \, \tilde{F}^{klmj} \, (\bar{\epsilon} \, \gamma_{klm} \psi) \, + \, a_7 \, \mu_6 \, ( {\bar\epsilon}^{*} \, \tilde{\gamma}^{j} \, \psi )] \, \Omega^{[5]}_{j} \nonumber\\
\delta \mathcal{L}_{\psi^{*}} \, & = & \, [- \, 5! \, i \, (h^{-1})^{ij} \, ({\bar\epsilon}^{*} \, \gamma_{i} \, \psi^{*}) \, + \,
(\mu_{1} \, \mathcal{F}^{ij} \, + \, \mu_{2} \, {\tilde{F \, F}}^{ij}) \, ({\bar\epsilon} \, \gamma_{i} \, \psi^{*}) \, + \, \nonumber\\
&& - (g_1 - g_2) \frac{a_7}{3} \, \tilde{F}^{klmj} \, (\bar{\epsilon}^{*} \, \gamma_{klm} \psi^{*}) \, + \, a_7 \, \mu_5 \, ({\bar\epsilon} \, \tilde{\gamma}^{j} \, \psi^{*} )] \, \Omega^{[5]}_{j}
\label{delta L}
\end{eqnarray}
where :
\begin{equation}
\begin{array}{rclcrcl}
\mu_{1} \, & = & \, - \, 4 \, 4! \, i \, a_3 \, (\Lambda^{1}_{+} \, + \, \Lambda^{2}_{+})
&;&
\mu_{2} \, & = & \, 4 \, [- 3 \, \nu \, i \, (\Lambda^{1}_{+} \, + \, \Lambda^{2}_{+}) \, - \, \frac{3}{16} \, g_1 \, a_7 \,
 (\Lambda^{1}_{+} \, - \, \Lambda^{2}_{+})] \\
\mu_{3} \, & = & \, - \, 4 \, 4! \, i \, a_3 \, (\Lambda^{1}_{-} \, + \, \Lambda^{2}_{-})
&;&
\mu_{4} \, & = & \, 4 \, [- 3 \, \nu \, i \, (\Lambda^{1}_{-} \, + \, \Lambda^{2}_{-}) \, - \, \frac{3}{16} \, g_1 \, a_7 \, (\Lambda^{1}_{-} \, - \, \Lambda^{2}_{-})]
\end{array}
\label{def mu}
\end{equation}
Recalling eq.s (\ref{Lambigcos}) and (\ref{Lfatcoset}) the above eq.s (\ref{def mu}) become:
\begin{equation}
\begin{array}{rclcrcl}
\mu_{1} \, & = & \, -5! \, i \, e^{\phi/2} &;&
\mu_{2} \, & = & \, \frac{3}{4} \, g_1 \, a_7 \, e^{-\frac{\phi}{2}} \\
\mu_{3} \, & = & \, -5! \, i \,  \, e^{\phi/2} &;&
\mu_{4} \, & = & \, - \, \frac{3}{4} \, g_1 \, a_7 \, e^{-\frac{\phi}{2}}\
\end{array}
\label{def mu cay}
\end{equation}
where we have chosen:
\begin{eqnarray}
\nu & = & \, \frac{g_1 \, a_7}{16} \, C_{[0]} \, = \, \frac{g_1 \, a_7}{16} \mbox{Re} \, \mathcal{N}
\label{val nu}
\end{eqnarray}
In the above equation we have introduced the complex kinetic matrix which would appear in a
$d=4$ gauge theory with scalars sitting in $\mathrm{SU(1,1)/U(1)}$ and
determined by the classical Gaillard--Zumino general formula \footnote{For a general discussion of the
Gaillard-Zumino formula see for instance \cite{parilez} } applied
to the specific coset:
\begin{equation}
  \mathcal{N} = \mbox{i}\,\frac{\Lambda^1_- - \Lambda^2_-}{\Lambda^1_- +
  \Lambda^2_-} \quad \Rightarrow \quad \cases{ \mbox{Re} \,
  \mathcal{N} =  C_0 \cr
  \mbox{Im} \, \mathcal{N} = e^{-\phi}\cr}
\label{scriptaNmata}
\end{equation}
It is convenient to rewrite the full variation (\ref{delta L}) of the Lagrangian in matrix form in the
$2$--dimensional space spanned by the fermion parameters $\left( \epsilon \, , \, \epsilon ^{*}\right) $ :
\begin{equation}
\delta \mathcal{L} \, = \, \delta \mathcal{L}_{\psi} +  \delta \mathcal{L}_{\psi^{*}} \,
= \, ( {\bar\epsilon} \, , \,  {\bar\epsilon}^{*})  \,
A  \, \,  \left( \begin{array}{c}
     \psi \\
    \psi^{*} \
 \end{array}\right)
\label{matr form}
\end{equation}
\begin{equation}
\hbox{{\tiny $ \! A_{p} =  \left( \begin{array}{ccc}
     - 5! \, i \, \gamma_{p} -20 \sqrt{2} a_7 \, \tilde{F}^{klmj} \, h_{jp} \, \gamma_{klm}  & &
       (-5! \, i \, e^{\frac{\phi}{2}} \mathcal{F}^{ij} \, \gamma_{i} \, - \, 30 \sqrt{2} \, a_7 \, e^{- \frac{\phi}{2}} {\tilde{F \, F}}^{ij} \, \gamma_i \, - \, \frac{2}{\sqrt{2}} \, a_7 \, e^{\frac{\phi}{2}} {\tilde \gamma}^{j}) \, h_{jp} \\
      (-5! \, i \, e^{\frac{\phi}{2}} \mathcal{F}^{ij} \gamma_i \, + \, 30 \sqrt{2} a_7 e^{-\frac{\phi}{2}}  {\tilde{F \, F}}^{ij} \gamma_i \, + \frac{2}{\sqrt{2}} a_7 \, e^{\frac{\phi}{2}} \, {\tilde \gamma}^{j})\, h_{jp} & &
       - 5! \, i \, \gamma_{p} + 20 \sqrt{2} \, a_7 \, {\tilde F}^{klmj} h_{jp}\gamma_{klm} \
 \end{array}\right)$}}
\label{matrix A}
\end{equation}
where $A \, = \, A_{p} \, \Omega_{[5]}^{p} $, and  $e^i \wedge e^j \wedge e^k \wedge e^l \wedge e^{m} \equiv \epsilon^{ijklmp}\Omega_{p}^{[5]}$ denotes the quadruplet of five--volume forms\\
The matrix $A_{p}$ is a tensor product of a matrices in spinor space and $2\times 2$ matrices
in the space spanned by $\left( \epsilon \, , \, \epsilon ^{*}\right)
$. It is convenient to spell out this tensor product structure which
is achieved by the following rewriting:
\begin{equation}
A_{p} \! = \! f_{1}  \gamma_{p} \otimes 1 \! \! 1 + f_{2} \, \tilde{F}^{klmj} \, h_{jp} \, \gamma_{klm} \otimes \sigma_{3} + f_{3} \, e^{\frac{\phi}{2}} \mathcal{F}^{ij} h_{jp} \, \gamma_i \otimes \sigma_{1} + (f_4 \,{\tilde \gamma}^{j} h_{jp} + f_5 \, e^{-\frac{\phi}{2}} \, {\tilde{F \, F}}^{ij} \gamma_i \, h_{jp}) \otimes \sigma_2
\label{concis Ak}
\end{equation}
where :
\begin{equation}
\begin{array}{rclcrclcrclcrclcrcl}
\!\!\! f_{1} \!\!\! &=& \!\!\! -5! \, i \, \rho \!\!\! &;& \!\!\! f_{2} \!\!\! &=& \!\!\! - 20 \sqrt{2} \, a_7   \!\!\! &;& \!\!\!
\! f_{3} \!\!\! &=& \!\!\! -5! \, i \, \rho \!\!\! &;& \!\!\! f_{4} \!\!\! &=& \!\!\! - \frac{2}{\sqrt{2}} \, i \, a_7\,  e^{\frac{\phi}{2}} \!\!\! &;& \!\!\! f_{5} \!\!\! &=& \!\!\! -30 \, i \, \sqrt{2} \, a_7
\end{array}
\label{gli f}
\end{equation}
and:
\begin{equation}
\begin{array}{rclcrclcrcl}
\tilde{\gamma} \, ^{i} & \equiv & \epsilon^{lmnpqi} \, \gamma_{lmnpq}
&;&
{\tilde F}^{klmp} & \equiv & \frac{1}{2} \, F_{ij} \, \epsilon^{ijklmp} &;&
({\tilde {F \, F}})_{ij} & \equiv & \frac{1}{2} \, \epsilon_{ijlmpq} \, F^{lm} \, F^{pq}
\end{array}
\label{pi 12}
\end{equation}
now using the solution of the first order equations already
determined in \cite{noidued3}, namely
\begin{equation}
 \left\{ \begin{array}{ccc}
 \widehat{G} \, = \, \eta \\
 \mathcal{F} \, = \frac{1}{\mu^{2}} h^{-1} F \, \eta \\
 h \eta \, = \, ( 1 \! \!  1 - \frac{1}{\mu^{2}} \, F \, \eta \, F \, \eta)
 \left[ \mbox{det} \left( \eta - \frac{1}{\mu} \, F \right) \right] ^{-1/2}
 \end{array}\right.
\label{sistem3}
\end{equation}
or, in more compact form:
\begin{equation}
h  \, = \, ( \eta  - \frac{1}{\mu^{2}} \, F^{2} )
\left[ \mbox{det} \left( \eta - \frac{1}{\mu} \, F \right) \right] ^{-1/2}
\label{h no eta}
\end{equation}
we can set
\begin{eqnarray}
 \frac{1}{\mu} & = & e^{-\phi/2} = \sqrt{\mbox{Im} \,
 \mathcal{N}}\nonumber\\
 \hat{F}  & \equiv & \sqrt{\mbox{Im} \,
 \mathcal{N}} F \nonumber \\
 \rho & = & e^{\frac{\phi}{2}}
\label{muchoi}
\end{eqnarray}
and we obtain:
\begin{eqnarray}
&& e^{\phi /2} \mathcal{F}^{ij} \, h_{jp} =
e^{\phi /2} e^{ - \phi}(Fh^{-1})^{ij} h_{jp} \equiv \hat{F}^{i} \, _{p} \nonumber \\
&& {\tilde F}^{klmp} = e^{\frac{\phi}{2}} \, {\tilde {\hat F}}^{klmp} \nonumber \\
&& e^{-\frac{\phi}{2}}({\tilde {F \, F}})_{ij} = e^{\frac{\phi}{2}} \, ({\tilde {\hat{F} \, \hat{F}}})_{ij}
\label{pi pi tilda}
\end{eqnarray}
This observation further simplifies the expression of $A_{k}$ which can be rewritten as:
\begin{equation}
A_{p} \! = \!e^{\frac{\phi}{2}}[z_{1} \gamma_{p} \otimes 1 \! \! 1 + z_{2} \, \tilde{\hat{F}}^{klmj} \, h_{jp} \, \gamma_{klm} \otimes \sigma_{3} + z_3 \, ({\tilde{\hat{F} \, \hat{F}}})^{ij} \gamma_i \, h_{jp} \otimes \sigma_2 + z_4 \hat{F}^{i}_{p} \, \gamma_i \otimes \sigma_1 + z_5 {\tilde \gamma}^{j} h_{jp} \otimes \sigma_2]
\label{concis2 Ak}
\end{equation}
where :
\begin{equation}
\begin{array}{rclcrclcrclcrclcrcl}
z_1 = f_{1} &;& z_2 = f_{2} &;& z_3 = f_5 &;& z_4 = f_3 &;& z_5 = f_4
\, .
\end{array}
\label{gli z}
\end{equation}
We introduce :
\begin{eqnarray}
&& \Pi^{p}_{(1)} = {\tilde {\hat{F}}}^{klmp} \, \gamma_{klm} \equiv \frac{1}{2} {\tilde \gamma }^{pij} \, F_{ij} \nonumber \\
&& \Pi^{p}_{(2)} = ({\tilde {\hat{F} \, \hat{F}}})^{ip} \, \gamma_i \nonumber \\
&& \Pi^{p}_{(3)} = F^{i} \, _{p} \, \gamma_i
\end{eqnarray}
and we obtain :
\begin{equation}
A_{p} \! = \! e^{\frac{\phi}{2}}[z_1 \, \gamma_{p} \otimes 1 \! \! 1 + z_{2} \Pi^{(1) j} \, h_{jp} \otimes \sigma_{3} + z_3 \Pi^{(2) j} \, h_{jp} + z_4 \Pi^{(3)}_{p} \otimes \sigma_1 + z_5 {\tilde \gamma}^{j} h_{jp} \otimes \sigma_2]
\label{A con Pi}
\end{equation}
Just as in the case of the $D3$--brane discussed in
\cite{noidued3}, the proof of $\kappa$--supersymmetry can now be reduced to the
following simple computation. Assume we have a matrix operator $\Gamma$
with the following properties:
\begin{eqnarray}
  \mbox{[a]}&\,& \Gamma^{2} = 1 \! \!  1 \nonumber\\
  \mbox{[b]}&\,& \Gamma \, A_{k} \, = \,  A_{k}
  \label{properte}
\end{eqnarray}
It follows that
\begin{equation}
P \, = \, \frac{1}{2} ( 1 \! \!  1 \, - \, \Gamma)
\label{proyector}
\end{equation}
is a projector since $P^2 = P$ and that
\begin{equation}
P A_{k} = \frac{1}{N}(1 \! \! 1 - \Gamma) \, A_{k} = 0
\label{k supersi pro}
\end{equation}
Therefore if we use supersymmetry parameters $\left( \overline{\kappa} , \overline{\kappa}^*\right)  =
\left (\overline{\epsilon} ,
\overline{\epsilon} ^*\right) \, P$ projected with this $P$, then the
action is invariant and this is just the proof of
$\kappa$--supersymmetry.
\par
The appropriate $ \Gamma $ is the following:
\begin{equation}
\Gamma \, = \, \frac{1}{N} [(\omega_{[6]} \, + \, \omega_{[2]}) \, \otimes \,
\sigma_{2} \, + \, (\omega_{[4]} \, + \, \omega_{[0]}) \, \otimes \, \sigma_{3}]
\label{def gamma}
\end{equation}
where:
\begin{eqnarray}
\omega_{[6]} \, & = & \, \alpha_{6} \, \epsilon^{ijklmn} \, \gamma_{ijklmn} \nonumber \\
\omega_{[4]} \, & = & \, \alpha_{4} \,  \epsilon^{ijklmn} \, \hat{F}_{ij} \, \gamma_{klmn} \nonumber \\
\omega_{[0]} \, & = & \, \alpha_{0} \,\epsilon^{ijklmn} \, \hat{F}_{ij} \, \hat{F}_{kl} \, \hat{F}_{mn} \nonumber\\
\omega_{[2]} \, & = & \, \alpha_{2} \, \epsilon^{ijklmn} \, \hat{F}_{ij} \, \hat{F}_{kl} \gamma_{mn} \nonumber \\
N \, & = & \, \left[ \mbox{det} \left( 1 \! \!  1 \, \pm \, \hat{F} \right) \right] ^{1/2}
\label{def omega}
\end{eqnarray}
and the coefficients are fixed to:
\begin{equation}
\begin{array}{rclcrclcrclcrcl}
\alpha_6 & = & \frac{i}{6!} &;&
\alpha_4 & = & \frac{1}{2 \, 4!} &;&
\alpha_0 & = & \frac{1}{48} &;&
\alpha_2 & = & \frac{i}{16}
\end{array}
\label{def alfa}
\end{equation}
This choice suffices to guarantee property $\mbox{[a]}$ in the above
list. An outline of the details of the calculation leading to such a result is given below. We have:
\begin{eqnarray}
\Gamma^{2} & = & \frac{1}{N^{2}} \left[\omega_{[6]}^{2} + \omega_{[2]}^{2} +
 \omega_{[6]} \, \omega_{[2]} + \omega_{[2]} \, \omega_{[6]} + \omega_{[4]}^{2} \omega_{[0]}^{2} +
2 \omega_{[4]} \omega_{[0]} \right] \otimes \, 1\!\!1_{2 \times 2} \nonumber \\
&& \hspace{0.3cm} + \left[\omega_{[6]}\omega_{[4]} - \omega_{[4]}\omega_{[6]} + \omega_{[6]} \omega_{[0]} - \omega_{[0]} \omega_{[6]} \right] \otimes \sigma_2 \, \sigma_3 = \nonumber \\
&& \hspace{-0.6cm} = \frac{1}{N^{2}} \left[\omega_{[6]}^{2} + \omega_{[2]}^{2} +
2 \omega_{[6]} \, \omega_{[2]} + \omega_{[4]}^{2} \omega_{[0]}^{2} +
2 \omega_{[4]} \omega_{[0]} \right] \otimes \, 1\!\!1_{2 \times 2}
\label{calc gamma}
\end{eqnarray}
and by straightforward manipulations we obtain
\begin{eqnarray}
\omega_{[6]}^{2} & = & - (\alpha_6)^{2}(6!)^{2} \nonumber \\
\omega_{[4]}^{2} & = & - (4!)^{2} (\alpha_{4})^{2} \,\hat{F}_{a_1 a_2} \,
\hat{F}_{a_3 a_4} \, \gamma^{a_1 a_2 a_3 a_4} - 2 (4!)^{2} (\alpha_{4})^{2} Tr(\hat{F}^{2}) \nonumber \\
\omega_{[2]}^{2} & = & \frac{4 (\alpha_2)^{2}}{9 \alpha_0 \alpha_4} \,
\omega_{[4]} \omega_{[0]} - 32 (\alpha_2)^{2} \left[Tr(\hat{F}^{2}) \, Tr(\hat{F}^{2}) -
2 \, Tr(\hat{F}^{4}) \right] \nonumber \\
\omega_{[6]} \omega_{[2]} & = & -2 \, (6!) \, \alpha_6 \, \alpha_2 \, \hat{F}_{a_1 a_2} \hat{F}_{a_3 a_4} \, \gamma^{a_1 a_2 a_3 a_4}
\end{eqnarray}
Using the following identities on determinants
\begin{equation}
\mbox{det} \left( 1 \!\! 1 \, \pm \, \hat{F} \right) = 1 \, - \, \frac{1}{2} Tr(\hat{F}^{2}) \,
+ \, \frac{1}{8} \left[Tr(\hat{F}^{2})\right]^{2} - \frac{1}{4} \, Tr(\hat{F}^{4}) \, + \mbox{det} \, (\hat{F})
\label{determinante eta}
\end{equation}
and
\begin{eqnarray}
\mbox{det} \, (\hat{F}) \, & = & \, - \, \frac{1}{6} \, Tr(\hat{F}^{6}) \, - \, \frac{1}{48} \, \left[Tr(\hat{F}^{2})\right]^{3} \, + \,
\frac{1}{8} \left[Tr(\hat{F}^{2})\right] \, \left[Tr(\hat{F}^{4})\right] = \\
& = & \left(\frac{1}{48} \, \epsilon_{a_1 a_2 a_3 a_4 a_5 a_6} \, \hat{F}^{a_1 a_2} \,
\hat{F}^{a_3 a_4} \, \hat{F}^{a_5 a_6} \right) \, =  \, \omega_{[0]}^{2} \\
\left(\hat{F} \, \tilde{\hat{F} \, \hat{F}}\right)_{ab}& = & - \,\frac{1}{12}\left( \epsilon_{a_1 a_2 a_3 a_4 a_5 a_6} \, \hat{F}^{a_1 a_2} \, \hat{F}^{a_3 a_4} \, \hat{F}^{a_5 a_6}\right)
\, 1 \! \! 1_{a b} \, = \, - 4 \, \omega_{[0]} \, 1 \! \! 1_{ab}
\label{prop 2 3}
\end{eqnarray}
that are just
consequences of $\widehat{F}=-\widehat{F}^T$ being antisymmetric
property $[a]$ is proved.
\par
Let us now turn to supersymmetry, namely to property $[b]$ :
\begin{equation}
\Gamma A_p \, = \, A_p
\label{torno susy}
\end{equation}
We need to calculate the products of the matrices $\omega_{[i]}$
$(i=6,4,2)$ defined above with the
gamma matrix structures appearing in the matrix $A_p$. Explicitly for
$\omega_{[6]}$ and omitting the simbol `` $ \hat{\phantom{.}} $ `` we find:
\begin{eqnarray}
\omega_{[6]} \, \gamma_{a} & = & \frac{i}{5!} \, {\tilde \gamma_{a}} \nonumber \\
\omega_{[6]} \, \Pi_{(1)}^{b_1} & = & \frac{3!}{2} \, i \, F_{b_2 b_3} \,
\gamma^{b_1 b_2 b_3} \nonumber \\
\omega_{[6]} \, \Pi_{(2)}^{b_2} & = & \frac{i}{5!} \, ({\tilde{F \, F}})^{b_1 b_2}
\, {\tilde \gamma_{b_1}} \nonumber \\
\omega_{[6]} \, \Pi_{(3)}^{b_2} & = & \frac{i}{5!} \, F^{b_1 b_2} \, {\tilde \gamma_{b_1}} \nonumber \\
\omega_{[6]} \, {\tilde \gamma^{b_1}} & = & -5! \, i \, \gamma^{b_1}
\label{susy om6}
\end{eqnarray}
while for $\omega_{[4]}$ we obtain
\begin{eqnarray}
\omega_{[4]} \, \gamma_{b_1} & = & \frac{1}{4!} \, {\tilde{F}}^{a_1 a_2 a_3 a_4} \,
\gamma_{a_1 a_2 a_3 a_4 b_1} - \frac{1}{3!} \, {\tilde{F}}_{b_1 a_2 a_3 a_4} \,
\gamma^{a_2 a_3 a_4} \nonumber \\
\omega_{[4]} \, \Pi_{(1)}^{b_4} & = & - \frac{3}{2} F_{a_3 a_4} \, F_{a_5 a_6} \,
\gamma^{b_4 a_3 a_4 a_5 a_6} \, -3 \, F^{b_5 b_6} \, F^{b_3 b_4} \,
\gamma_{b_5 b_6 b_3} \, -3 \, Tr(F^{2}) \gamma^{b_4} \, +6 \, (F^2)^{b_5} \, _{b_6}
\, \gamma^{b_6} \nonumber \\
\omega_{[4]} \, \Pi_{(2)}^{b_2} & = & \frac{1}{4!} {\tilde F}_{a_1 a_2 a_3 a_4} \,
({\tilde {F \, F}})_{b_1} \, ^{b_2} \, \gamma ^{a_1 a_2 a_3 a_4 b_1} \, - \,
\frac{1}{6} \, {\tilde F}_{a_1 a_2 a_3 a_4} \, ({\tilde {F \, F}})^{a_1 b_2} \,
\gamma^{a_2 a_3 a_4} \nonumber \\
\omega_{[4]} \Pi_{(3)}^{b_2} & = & \frac{1}{4!} \, {\tilde F}_{a_1 a_2 a_3 a_4}
\, F_{b_1} \, ^{b_2} \, \gamma^{a_1 a_2 a_3 a_4 b_1} \, + \,
\frac{1}{12} F^{a_5 a_6} F^{a_1 b_2} \, \gamma_{a_1 a_5 a_6}\nonumber \\
\omega_{[4]} {\tilde \gamma^{b_6}} & = & \frac{5!}{2} \, F_{b_4 b_5} \,
\gamma^{b_6 b_4 b_5} \, + \, 5! \, F^{b_5 b_6} \, \gamma_{b_5}
\label{susy om4}
\end{eqnarray}
and finally for $\omega_{[2]}$ we have:
\begin{eqnarray}
\omega_{[2]} \, \gamma_{b_1} & = & \frac{i}{8} \, ({\tilde{F \, F}})_{a_1 a_2} \,
\gamma^{a_1 a_2}_{b_1} \, - \, \frac{i}{4} \, ({\tilde{F \, F}})_{b_1 a_2} \,
\gamma^{a_2} \nonumber \\
\omega_{[2]} \, \Pi_{(1)}^{b_4} & = & \frac{i}{8}  ({\tilde{F F}})_{a_1 a_2} \,
 {\tilde F}^{b_1 b_2 b_3 b_4} \, \gamma^{a_1 a_2}_{b_1 b_2 b_3}  -
\frac{3}{4} \, i \, ({\tilde{F F}})_{a_1 a_2} \, {\tilde F}^{a_1 b_2 b_3 b_4} \,
\gamma^{a_2}_{b_2 b_3} - \frac{3}{4} \, i \, ({\tilde{F F}})_{a_1 a_2} \,
{\tilde F}^{a_1 a_2 b_3 b_4} \, \gamma_{b_3} \nonumber \\
\omega_{[2]} \, \Pi_{(2)}^{b_2} & = & \frac{i}{8} ({\tilde{F F}})_{a_1 a_2} \,
 F^{b_1 b_2} \, \gamma^{a_1 a_2}_{b_1} \, - \, \frac{i}{4} \, ({\tilde{F F}})_{b_1 a_2} \,
  ({\tilde{F F}})^{b_1 b_2} \, \gamma^{a_2} \nonumber \\
\omega_{[2]} \, \Pi_{(3)}^{b_2} & = & \frac{i}{8} \, ({\tilde{F \, F}})_{a_1 a_2} \,
 F^{b_1 b_2} \, \gamma^{a_1 a_2}_{b_1} + \frac{1}{4} [F \, ({\tilde{F F}})]^{b_2} \, _{a_2} \,
 \gamma^{a_2} \nonumber \\
\omega_{[2]} \, {\tilde \gamma^{b_6}} & = & - \frac{5}{2} \, i \, ({\tilde{F \, F}})_{a_1 a_2}
\, {\tilde \gamma^{a_1 a_2 b_6}}
\label{susy om2}
\end{eqnarray}
With these relations we determine the values of the parameter
$a_7 \, = \, \frac{\sqrt{2}}{2} \, i$ from the supersymmetry relation [b]
 (\ref{properte}) and the projector (\ref{def gamma}).
\section{Outlook and conclusions}
\label{outlook}
In this paper we have applied the new first order formalism for
$p$--brane world volume actions introduced by us in \cite{noidued3} to the case of
$5$ branes. The motivations for such a calculation were already discussed in the introduction and
are not repeated here. We recall the essential features of the new formalism  that
we have adopted and that allows to reproduce the
Born--Infeld second order action via the elimination of a set composed by three auxiliary fields:
\begin{itemize}
  \item $\Pi^{\underline{a}}_i$
  \item $h^{ij}$ (symmetric)
  \item $\mathcal{F}^{ij}$ (antisymmetric)
\end{itemize}
Distinctive properties are:
\begin{enumerate}
  \item All fermion fields are implicitly hidden inside
  the definition of the $p$--form potentials of supergravity
  \item $\kappa$--supersymmetry is easily proven from supergravity
  rheonomic parametrizations
  \item The action is manifestly covariant with respect to the duality group
  $\mathrm{SL(2,\mathbb{R})}$ of type IIB supergravity.
  \item The action functional can be computed on any background which
  is an exact solution of the supergravity bulk equations.
\end{enumerate}
Of specific interest in applications are precisely the last two
properties. Putting together our result we can summarize the $D5$ brane action we have found as follows:
\begin{eqnarray}
\mathcal{L} & = & \frac{1}{\sqrt{\mbox{Im}\mathcal{N}}}  \Pi^{\underline{a}} _i \, V^{\underline{b}} \,
\eta_{\underline{ab}}
 \, \eta^{i\ell_1} \, \wedge e^{\ell_2} \, \wedge \, \dots \wedge e^{\ell_6} \, \epsilon_{\ell_1 \dots
 \ell_6}  -  \frac{1}{12 \, \sqrt{\mbox{Im}\mathcal{N}}}  \Pi^{\underline{a}} _i \, \Pi^{\underline{b}}
_j \, \eta_{\underline{ab}} \, h^{ij} \, e^{\ell_1} \, \wedge \, \dots
\wedge e^{\ell_6} \, \epsilon_{\ell_1 \dots \ell_6} \nonumber\\
&& - \, \frac{1}{3 \, \sqrt{\mbox{Im}\mathcal{N}}} \, \left [ \mbox{det}\,\left(  h^{-1}
+ \frac{1}{\sqrt{\mbox{Im}\mathcal{N}}} \, \mathcal{F}\right) \right] ^{\frac{1}{4}} \,
  e^{\ell_1} \, \wedge \, \dots
\wedge e^{\ell_6} \, \epsilon_{\ell_1 \dots \ell_6}\nonumber\\
&& + \frac{5}{4 \, \sqrt{\mbox{Im}\mathcal{N}}} \, \mathcal{F}^{ij} \, F^{[2]} \,
\wedge \, e^{\ell_3} \,\wedge \dots  \wedge \,e^{\ell_6} \,
\epsilon_{ij\ell_3 \dots \ell_6} \nonumber \\
&& - \, \frac{5}{2} \, i \, \mbox{Re} \mathcal{N} \, F \, \wedge \, F \, \wedge \, F + \frac{5}{2} \, i
\,  p_{\beta}  \, A^{\beta} \,
\wedge \, F \wedge \, F + 40 \, i \, C_{[4]} \wedge \, F + \frac{\sqrt{2}}{2} \, i \,
q_{\alpha} C_{[6]}^{\alpha} 
\label{D3branaczia}
\end{eqnarray}
where:
\begin{eqnarray}
F & \equiv & F_{[G]} + q_{\alpha} A^{\alpha}\nonumber\\
\mathcal{N} &=& \,\frac{p_{\alpha} \, \Lambda^{\alpha}_- }{q_{\beta} \, \Lambda^{\beta}_-} \nonumber \\
 A^{\alpha} & = & \frac{1}{\sqrt{2}} \, 
( B_{[2]} - i C_{[2]} \, , \, B_{[2]} + i C_{[2]}) \nonumber \\
 q_{\alpha} & = & \frac{1}{\sqrt{2}} (p+iq \, , \,  p-iq) \nonumber \\
p_{\alpha}& = & \frac{1}{\sqrt{2}} (r+is \, , \, r-is) \nonumber \\
1 & = & q_{\alpha} \, p_{\beta} \, \epsilon^{\alpha \beta}
\end{eqnarray}
In the particular case of $q_\Lambda =\{1,0\}$ that implies $q_\alpha
= \frac{1}{\sqrt{2}}(1, 1)$ and 
$p_\Lambda =\{0,1\}$ that implies $p_\alpha
= \frac{i}{\sqrt{2}}(1, -1)$ 
 we have:
\begin{equation}
\mathcal{N} = \mbox{i}\,\frac{\Lambda^1_- - \Lambda^2_-}{\Lambda^1_- +
  \Lambda^2_-} \, = \, i \, e^{- \phi} \, + \, C_{[0]}
\end{equation}
As many times stressed, our specific interest in the above action is
given by its evaluation on the background provided
by the bulk solution found in \cite{noialtrilast} which describes a $D3$--brane with an
$\mathbb{R}^2 \times ALE$ transverse manifold and a $2$--form flux trapped on
a homology $2$--cycle of the ALE.  Combining this with the world volume action of
a $D3$--brane obtained in \cite{noidued3} we will finally obtain the appropriate
source term of that exact solution
which was so far missing. Alternatively by expanding (\ref{D3branaczia}) for small fluctuations around
the same background we can use it as a token to explore
the gauge/gravity correspondence. The key point is to reconcile this
with the fact there is no running dilaton  in the bulk  solution of
\cite{noidued3}. This will achieved by an appropriate choice of the
charge vectors $q_\Lambda$ and $p_\Lambda$, thanks to present manifestly
$SL(2,R)$-covariant formulation. Such applications are postponed to a
forthcoming paper.
\section*{Acknowledgements}
We are grateful to Marco Bill\'o, Emiliano Imeroni, Alberto Lerda, Igor Pesando, and Antonio Rago for many important and clarifying discussions about this work.
\par
\appendix
\section{Rheonomic parametrization of the type IIB supergravity
curvatures}
In order to obtain the supersymmetry transformation rules used in the
text one needs the rheonomic parametrizations of the curvatures.
For simplicity we write them only in the complex basis and we disregard the bilinear
fermionic terms calculated by Castellani and Pesando. We have:
\begin{eqnarray}
 R^{\underline{a}}&=&0\nonumber\\&{}\label{torpar}\\
 \rho&=&\rho_{\underline{ab}} V^{\underline{a}} \wedge V^{\underline{b}}
      + {5\over 16}{\rm i}\Gamma^{\underline{a_1-a_4}} \psi V^{\underline{a_5}} \left(
          F_{\underline{a_1-a_5}}
          +{1\over5!}\epsilon_{\underline{a_1-a_{10}}}F_{\underline{a_6-a_{10}}}
        \right)\nonumber\\
     & &+{1\over 32}\left(
         -\Gamma^{\underline{a_1-a_4}}\psi^*V_{\underline{a_1}}
         +9\Gamma^{\underline{a_2a_3}}\psi^*V^{\underline{a_4}}
       \right)\Lambda^{\alpha}_+\mathcal{H}^{\beta}_{\underline{a_2-a_4}} \epsilon_{\alpha\beta}\nonumber\\
     & &+\mbox{fermion bilinears}\label{rhopar}\\
&{}&\nonumber\\
 \mathcal{H}^{\alpha}_{[3]}&=&
     \mathcal{H}^{\alpha}_{\underline{abc}} V^{\underline{a}} \wedge V^{\underline{b}} \wedge
     V^{\underline{c}}
    +\Lambda^{\alpha}_+ {\bar \psi}^* \Gamma_{\underline{ab}}\lambda^* V^{\underline{a}} \wedge
    V^{\underline{b}}
    +\Lambda^{\alpha}_- {\bar \psi} \Gamma_{\underline{ab}} \lambda V^{\underline{a}} \wedge V^{\underline{b}}
    \label{Hpar}\\
&{}&\nonumber\\
\mathcal{F}_{[5]}&=&F_{\underline{a_1-a_5}}V^{\underline{a_1}}
\wedge \dots \wedge V^{\underline{a_5}}\label{f5par}\\&{}\nonumber\\
{\cal D} \lambda&=&
   {\cal D}_{\underline{a}}\lambda V^{\underline{a}}
   +{\rm i} P_{\underline{a}} \Gamma^{\underline{a}} \psi^*
   -{1 \over 8}{\rm i}\Gamma^{\underline{a_1-a_3}} \psi
      \epsilon_{\alpha\beta}\Lambda^{\alpha}_+\mathcal{H}^{\beta}_{\underline{a_1-a_3}}
      \label{dilatinpar}\\&{}&\nonumber\\
{\cal D}\Lambda^{\alpha}_+&=&
    \Lambda^{\alpha}_- P_{\underline{a}}V^{\underline{a}}
    + \Lambda^{\alpha}_-{\bar \psi}^* \lambda
\label{GsuHpar1}\\&{}&\nonumber\\
{\cal D}\Lambda^{\alpha}_-&=&
   \Lambda^{\alpha}_+P_{\underline{a}}^*V^{\underline{a}}
   + \Lambda^{\alpha}_+{\bar \psi}\lambda^*
\label{GsuHpar2}\\
R^{\underline{ab}}&=&R^{\underline{ab}}_{~~\underline{cd}} V^{\underline{c}} \wedge V^{\underline{d}} \,
+ \, \mbox{fermionic
terms}\label{lorenpar}\\
\mathcal{H}^{\alpha}_{[7]}&=& \frac{1}{3! \, 7!} \,
     \epsilon^{\underline{abcd_1\dots d_7}}\mathcal{H}^{\alpha}_{\underline{abc}}
     V_{\underline{d_1}} \wedge \dots \wedge
     V_{\underline{d_7}}
    +\dots
    \label{H7par}\\
&{}&\nonumber\\
\end{eqnarray}
\section{$\mbox{SL}(2,\mathbb{R})$ and $\mbox{SU}(1,1)$ covariant formalism for D-branes}
\label{covariance}
We define a two component vector for the fields $B_{[2]} $ and $C_{[2]}$:
\begin{equation}
 A^{\Sigma} = (B_{[2]} , C_{[2]})
 \label{Avec}
\end{equation}
The variation of this vector and $\mathcal{N}$ under $SL(2,R)$ are :
\begin{equation}
 \left( \begin{array}{c}
B_{[2]}^{\prime} \\
C_{[2]}^{\prime} 
\end{array}\right) = 
\left( \begin{array}{cc}
    p  &  q \\
    r  &  s
  \end{array}\right)
\left( \begin{array}{c}
B_{[2]} \\
C_{[2]} 
\end{array}\right) = 
\left( \begin{array}{c}
p \, B_{[2]} + q \, C_{[2]} \\
r \, B_{[2]} + s \, C_{[2]}
\end{array}\right)
\end{equation}
\begin{equation}
\mathcal{N}^{\prime} = \frac{s \, \mathcal{N} + r}{q \, \mathcal{N} + p} = 
\frac{s q \, (C_{[0]}^{2} + e^{-2 \phi}) + (s p + r q) \, C_{[0]} + r p + 
i \, e^{- \phi}}{q^{2} \, (C_{[0]}^{2} + e^{-2 \phi}) + 2 q p \, C_{[0]} + p^{2}}
\end{equation}
We can see that there is a subgroup of $\mbox{SL}(2,\mathbb{R})$ that leaves invariant the Lagrangian. \\
If we take the element :
\begin{equation}
  T  = \left( \begin{array}{cc}
    1     &     0 \\
    r    &     1
  \end{array}\right)
  \label{Tmatrix}
\end{equation}
we obtain for D3-brane (we have the same results for D5-brane) :
\begin{eqnarray}
 && B_{[2]}^{\prime} = B_{[2]} \,\,\, , \,\,\, C_{[2]}^{\prime} = r B_{[2]} + C_{[2]} \nonumber \\
 && \mathcal{N}^{\prime} = \mathcal{N}+ r \nonumber \\
 && Im(\mathcal{N}^{\prime}) = Im(\mathcal{N}) \,\,\, , \,\,\, Re(\mathcal{N}^{\prime}) = Re(\mathcal{N}) + r  
 \end{eqnarray}
\begin{eqnarray}
&& \delta_{T} \mathcal{L}_{B.I} = 0 \nonumber \\
&& \mathcal{L}_{W.Z.T}^{\prime} = - \ft 3 4 i \, [\mbox{Re} \, \mathcal{N} + r] \, \mbox{}\, B_{[2]} \, \wedge \, B_{[2]} + \ft 3 4 \, i \, [C_{[2]} + r B_{[2]}] \,
\wedge \, B_{[2]} + 6 i \, C_{[4]} = \mathcal{L}_{W.Z.T}
\label{invarianza}
\end{eqnarray}
this is a one-parameter subgroup of $\mbox{SL}(2,\mathbb{R})$ \\
Now we introduce e covariant formalism for $\mbox{SL}(2,\mathbb{R})$ covariance; in particular we introduce two vectors $q_{\Lambda}$ and $p_{\Lambda}$ that transform in the fundamental representation of $\mbox{SL}(2,\mathbb{R})$. In analogy with $\varphi_{i} \, \hat{n}_{i}$ in field theory, where $\hat{n}_{i}$ is a fix vector and $\varphi_{i}$ is a scalar field in the fundamental representation of $\mbox{SO(N)}$, to do this we replace in the Lagrangian :
\begin{eqnarray}
&& B_{[2]} \rightarrow q_{\Lambda} A^{\Lambda} \nonumber \\
&& C_{[2]} \rightarrow p_{\Lambda} A^{\Lambda} 
 \end{eqnarray} 
where $A^{\Lambda}$ is definite in (\ref{Avec}).\\
The vectors $q_{\Lambda}$ and $p_{\Lambda}$  have this general form :
\begin{eqnarray}
&& q_{\Lambda} = (p , q) \nonumber \\
&& p_{\Lambda} = (r , s)
\label{PiQu} 
 \end{eqnarray} 
with the constrain :
\begin{equation}
q_{\Lambda} p_{\Sigma} \epsilon^{\Lambda \Sigma} = 1
\label{constrein}
\end{equation}
This is actually the condition that the determinant of a general element of  $\mbox{SL}(2,\mathbb{R})$  is equal to one.\\
We have in total only three real parameters but as we will see in the $\mbox{SU}(1,1)$ formalism one of those is connected to the invariance (\ref{invarianza}).\\
We must introduce also the covariant form of $\mathcal{N}$, but, to do this, is suitable to pass at the  $\mbox{SU}(1,1)$ formalism.
We pass to $\mbox{SU}(1,1)$ formalism using the Cayley matrix :
\begin{equation}
  C^{\alpha} \, _{\Lambda}  = \frac{1}{\sqrt{2}}  \left( \begin{array}{cc}
    1     &     - i \\
    1     &     i
  \end{array}\right)
\end{equation}
and the inverse matrix :
\begin{equation}
\tilde{C}_{\alpha} \, ^{\Lambda} = (C^{-1}) ^{\Lambda} \, _{\alpha}  = \frac{1}{\sqrt{2}}  \left( \begin{array}{cc}
    1     &     1 \\
    i    &      -i
  \end{array}\right)
\end{equation}
In this way we have that :
\begin{eqnarray}
&& q_{\alpha} = \tilde{C}_{\alpha} \, ^{\Lambda} \, q_{\Lambda} \nonumber \\
&& p_{\alpha} = \tilde{C}_{\alpha} \, ^{\Lambda} \, p_{\Lambda} \nonumber \\
&& A^{\alpha} = C^{\alpha} \, _{\Lambda} \, A^{\Lambda} \nonumber \\
\label{PiQuS11} 
 \end{eqnarray} 
and :
\begin{equation}
q_{\alpha} \, A^{\alpha} = \tilde{C}_{\alpha} \, ^{\Sigma} \, q_{\Sigma} \, C^{\alpha} \, _{\Lambda} \, A^{\Lambda} = (C^{-1}) ^{\Sigma} \, _{\alpha} \, C^{\alpha} \, _{\Lambda} \, A^{\Lambda} = q_{\Lambda} \, A^{\Lambda} 
\end{equation}
From (\ref{Avec}), (\ref{PiQu}), (\ref{PiQuS11})
\begin{eqnarray}
&& A^{\alpha} = C^{\alpha} \,\, _{\Sigma} \, A^{\Sigma}  = \frac{1}{\sqrt{2}} \, 
( B_{[2]} - i C_{[2]} \, , \, B_{[2]} + i C_{[2]}) \nonumber \\
&& q_{\alpha} = \frac{1}{\sqrt{2}} (p+iq \, , \,  p-iq) = \frac{1}{\sqrt{2}} (q_1 , q_1^{\star}) \nonumber \\
&& p_{\alpha} = \frac{1}{\sqrt{2}} (r+is \, , \, r-is) = \frac{1}{\sqrt{2}} (p_1 , p_1^{\star})
\label{qp}
 \end{eqnarray}
The vectors $q_{\Lambda}$ and $p_{\Lambda}$ transform in the fundamental representation of $\mbox{SL(2,R)}$ as :
\begin{eqnarray}
&& q_{\Lambda}^{\prime} = q_{\Sigma} \, U^{\Sigma}\,_{\Lambda} \nonumber \\
&& p_{\Lambda}^{\prime} = p_{\Sigma} \, U^{\Sigma}\,_{\Lambda} 
\end{eqnarray}
and in the formalism of $\mbox{SU(1,1)}$ the vectors $q_{\alpha}$ and $p_{\alpha}$ transform as :
\begin{eqnarray}
q_{\alpha}^{\prime} = q_{\Sigma}^{\prime} \, (C^{-1})^{\Sigma}\,_{\alpha} = q_{\Lambda} \, U^{\Lambda}\,_{\Sigma}  \, (C^{-1})^{\Sigma}\,_{\alpha} = q_{\beta} \, C^{\beta}\,_{\Lambda} \, U^{\Lambda}\,_{\Sigma}  \, (C^{-1})^{\Sigma}\,_{\alpha}\,
\end{eqnarray}
so the relation between $\mbox{SU(1,1)}$ and $\mbox{SL(2,R)}$ is :
\begin{equation}
U^{ \beta}\,_{\alpha} = C^{\beta}\,_{\Lambda} \, U^{\Lambda}\,_{\Sigma}  \, (C^{-1})^{\Sigma}\,_{\alpha} = (C \, U \, C^{-1})^{\beta}\,_{\alpha}
\end{equation}
\begin{equation}
 U^{\alpha}\,_{\beta} =  \frac{1}{2} \left( \begin{array}{cc}
    1     &     -i \\
     1     &     i
  \end{array}\right)
 \left( \begin{array}{cc}
    p     &     q \\
    r     &       s
  \end{array}\right)
  \left( \begin{array}{cc}
    1     &     1 \\
     i     &     -i
  \end{array}\right) = \frac{1}{2}
  \left( \begin{array}{cc}
    p -ir + iq +s     &     p-ir-iq-s \\
    p+ir+iq-s    &          p+ir-iq+s
  \end{array}\right)
  \label{matrixSU}
  \end{equation}
The property (\ref{qp}) is invariant under $\mbox{SU(1,1)}$, in fact :
\begin{eqnarray}
&& q^{\prime}_{\alpha} = q_{\beta} \, U^{ \beta}\,_{\alpha} \nonumber \\
&& q^{\prime \, \star}_{\alpha} = q_{\beta}^{\star} \, \left(U^{ \beta}\,_{\alpha}\right)^{\star}\nonumber \\
&& q^{\prime \, \star}_1 = q^{\star}_1 \, \left(U^{1}\,_1\right)^{\star} + q^{\star}_2 \, \left(U^{2}\,_1\right)^{\star}  = q^{\star}_1 \, \left(U^{1}\,_1\right)^{\star} + q_1 \, \left(U^{2}\,_1\right)^{\star}  
\end{eqnarray}
on the other side we have that 
$q^{\prime}_2 = q_1 \, \left(U^{1}\,_2\right) + q_2 \, \left(U^{2}\,_2 \right) = q_1 \, \left(U^{1}\,_2\right) + q_1^{\star} \, \left(U^{2}\,_2 \right) $
and so if we want that $q_2^{\prime} = q^{\prime \star}_1$ we must have that $\left(U^{2}\,_1\right)^{\star}  = U^{1}\,_2$ and $\left(U^{1}\,_1\right)^{\star} = U^{2}\,_2$, but this is exactly the property of SU(1,1) how we can see in (\ref{matrixSU}).\\
The Levi-Civita tensor becomes :
\begin{equation}
  \\\epsilon^{\alpha \beta} = C^{\alpha} \,\,_{\Lambda} \, C^{\beta} \,\,_{\Sigma} \, \epsilon^{\Lambda \Sigma}  = \left( \begin{array}{cc}
    0     &     i \\
    - i     &     0
  \end{array}\right)
\end{equation}
and it is very simple to verify that the constrain now is :
\begin{equation}
q_{\alpha} \, p_{\beta} \, \epsilon^{\alpha \beta} = 1
\label{determ}
\end{equation}
The covariant formulation of the $\mathcal{N}$ matrix is:
\begin{equation}
\mathcal{N} = \, i \, \frac{\Lambda^1_- - \Lambda^2_-}{\Lambda^1_- +
  \Lambda^2_-} \, = \, i \, e^{- \phi} \, + \, C_{[0]}  \,\, \rightarrow \,\, 
\frac{p_{\alpha} \Lambda^{\alpha}_{-}}{q_{\beta} \Lambda^{\beta}_{-}}
\end{equation}
The condition (\ref{determ}) says that 
\begin{equation}
ps - qr = 1
\label{determ.esp}
\end{equation}

{\bf Now we can see some examples}.\\
$\bullet$ If $p \neq 0$\\
\begin{equation}
s = \frac{1}{p} + \frac{qr}{p}
\end{equation}
and so 
\begin{equation}
p_{\alpha} = \frac{1}{\sqrt{2}} \left( r + i \, \frac{1}{p} + i \,\frac{qr}{p} \, , r - i \, \frac{1}{p} - i \,\frac{qr}{p} \right) = \frac{i}{p \, \sqrt{2}} (1 \, , \, -1) \, + \, \frac{r}{p} \, q_{\alpha} \equiv \tilde{p}_{\alpha} \, + \, \frac{r}{p} \, q_{\alpha}  
\end{equation}
The $\mathcal{N}$ matrix became 
\begin{equation}
\mathcal{N}  = \frac{r}{p} + 
\frac{\tilde{p}_{\alpha} \Lambda^{\alpha}_{-}}{q_{\beta} \Lambda^{\beta}_{-}}
\end{equation}
and it is very simple to see that the parameter $r$ disapperaes from the Lagrangian, infact it is the parameter of the invariance (\ref{Tmatrix}). How we said before only two parameters ($ p$ and $q$ in this case) are independent and produce $\mbox{SL}(2,\mathbb{R})$ covariant Lagrangianes.\\
$\bullet$ If $p = 0 $
\begin{eqnarray}
&& q_{\alpha} = \frac{i \, q}{\sqrt{2}} (1 \, , \, -1) \label{qdual}\\
&& p_{\alpha} = \frac{1}{\sqrt{2}} (r + is \, , \, r -is)
\end{eqnarray}
from (\ref{determ}) or (\ref{determ.esp}) 
\begin{eqnarray}
 r = - \frac{1}{q}
\end{eqnarray}
and 
\begin{equation}
p_{\alpha} = \frac{1}{\sqrt{2}} \left( - \frac{1}{q} + i \, s \, , - \frac{1}{q} - i \, s \right) = \frac{i \, s}{\sqrt{2}} (1 \, , \, -1) \, - \, \frac{1}{q \, \sqrt{2}} \, (1 \, , \, 1) \equiv \frac{s}{q} \, q_{\alpha}  + \tilde{p}_{\alpha}
\label{pdual}
\end{equation}
The $\mathcal{N}$ matrix now is 
\begin{equation}
\mathcal{N}  = \frac{s}{q} + 
\frac{\tilde{p}_{\alpha} \Lambda^{\alpha}_{-}}{q_{\beta} \Lambda^{\beta}_{-}} = \frac{s}{q} - \frac{1}{q^{2} \, \mathcal{N}} 
\end{equation}
and 
\begin{eqnarray}
 && q_{\alpha} \, A^{\alpha} \, = \, q \, C_{[2]} \nonumber \\
 && \tilde{p}_{\alpha} \, A^{\alpha} \, = \, - \frac{1}{q} \, B_{[2]}  
\end{eqnarray}
In this case the invariance parameter is $s$, the only independent parameter is $q$ and the Lagrangian is that we can obtain with the duality:
\begin{equation}
  D  = \left( \begin{array}{cc}
    0     &     1 \\
    -1    &      0
  \end{array}\right)   
  \label{Dmatrix}
\end{equation}
At this point we can ask to ourself what is the relaction between $p \neq 0$ and $p=0$. If we start 
from the SL(2,R) vectors $q_{\Lambda} = (1 , 0)$ and $p_{\Lambda} = (0 ,1)$ we can see that the stability group of $q_{\Lambda}$ is $T$ of (\ref{Tmatrix}) :
\begin{eqnarray}
&& q_{\Lambda}^{\prime} = (1,0) \left( \begin{array}{cc}
    1     &     0 \\
    r    &    1
  \end{array}\right) = (1,0)  \nonumber \\
&&   p_{\Lambda}^{\prime} = (0,1) \left( \begin{array}{cc}
    1     &     0 \\
    r    &    1
  \end{array}\right) = r \, (1,0) + (0,1) = p_{\Lambda} + r \, q_{\Lambda}
  \end{eqnarray}
Now we transform $q_{\Lambda}$ and $p_{\Lambda}$ with the duality 
$U = D$ of (\ref{Dmatrix}) and obtain $q_{\Lambda}^{\prime} = (0,1)$ and $p_{\Lambda}^{\prime} = (-1,0)$, contemporary :
\begin{eqnarray}
T ^{\Lambda}\,_{\Sigma}  \rightarrow \left(U^{-1} T \, U\right)^{\Lambda}\,_{\Sigma} = 
 \left( \begin{array}{cc}
    1     &    -r \\
    0    &    1
  \end{array}\right)
   \end{eqnarray}
this transformations are the stability group of $q^{\prime}_{\Lambda} = (0,1)$, in fact :
\begin{eqnarray}
&& q^{\prime}_{\Lambda} = (0,1) \left( \begin{array}{cc}
    1     &    -r \\
    0    &    1
  \end{array}\right) = 
  (0,1)\nonumber \\
  && p^{\prime}_{\Lambda} = (-1,0) 
   \left( \begin{array}{cc}
    1     &    -r \\
    0    &    1
  \end{array}\right) = 
(-1,0) + r \, (0,1) = p_{\Lambda} + r \, q_{\Lambda}  
\label{qp.duali}
 \end{eqnarray}
but as you can see in (\ref{casoduale}) $q^{\prime}_{\Lambda} = (0,1) \rightarrow q_{\alpha} = \frac{i}{\sqrt{2}}(1, -1)$ and $p^{\prime}_{\Lambda} = (-1,0) \rightarrow p_{\alpha} = -\frac{1}{\sqrt{2}}(1,1)$ when we pass to SU(1,1) formalism, 
 and so \ref{qp.duali} become exactly the relations (\ref{qdual}), (\ref{pdual}) if we take $q=1$.
 \subsection{Particular choice of $q_{\alpha}$ and $p_{\alpha}$}
$\diamond$ For :
\begin{eqnarray}
&& q_{\Lambda} = (1 , 0) \,\,\, \Rightarrow  q_{\alpha} = q_{\Lambda}(C^{-1})^{\Lambda} \,\,_{\alpha} = \frac{1}{\sqrt{2}} (1 , 1) \nonumber \\
&& p_{\Lambda} = (0 , 1) \,\,\, \Rightarrow  p_{\alpha} = p_{\Lambda} \, (C^{-1})^{\Lambda} \,\,_{\alpha}  = \frac{i}{\sqrt{2}}(1 , -1) 
\nonumber
\end{eqnarray}
we have :
\begin{eqnarray}
&& q_{\alpha} A^{\alpha} = B_{[2]} \nonumber \\
&& p_{\alpha} A^{\alpha} = C_{[2]} \nonumber \\
&& q_{\alpha} p_{\beta} \epsilon^{\alpha \beta} = 1
\end{eqnarray}
$\diamond$ For :
\begin{eqnarray}
&& q_{\Lambda} = (0 , 1) \,\,\, \Rightarrow  q_{\alpha} =  \frac{i}{\sqrt{2}}(1\, , \, -1)  \nonumber \\
&& p_{\Lambda} = (-1 , 0) \,\,\, \Rightarrow p_{\alpha} = - \frac{1}{\sqrt{2}}(1\, , \, 1) 
\label{casoduale}
\end{eqnarray}
we have :
\begin{eqnarray}
&& q_{\alpha} A^{\alpha} = C_{[2]} \nonumber \\
&& p_{\alpha} A^{\alpha} = - B_{[2]}
\end{eqnarray}
\begin{equation}
\mathcal{N} \, \rightarrow \, \mathcal{N}_{Dual} = - \, \frac{1}{\mathcal{N}}
\end{equation}
and 
\begin{eqnarray}
Im(\mathcal{N}_{Dual}) = \frac{1}{e^{-\phi} + C_{[0]}^{2} \, e^{-\phi}} \nonumber \\
Re(\mathcal{N}_{Dual}) = - \frac{e^{\phi} \, C_{[0]}}{e^{-\phi} + C_{[0]}^{2} \, e^{\phi}} 
\end{eqnarray}
\subsection{Invariant Tensors under $\mbox{SU}(1,1)$} 
The invariant tensors under $\mbox{SU}(1,1)$ are the metric $\eta^{\alpha \beta}$ and the tensor $\epsilon^{\alpha \beta}$, in fact :
\begin{eqnarray}
&& U^{T} \epsilon \, U = \epsilon \\
&&U^{\dagger} \eta \, U = \eta\\
\end{eqnarray}
where we defined :
\begin{equation}
  \eta^{\alpha \beta} = \left( \begin{array}{cc}
    1     &     0 \\
    0    &    -1
  \end{array}\right) \hspace{2cm}
  \epsilon^{\alpha \beta} = \left( \begin{array}{cc}
    0     &     i \\
    -i    &      0
  \end{array}\right)
    \label{eta.matrix}
\end{equation}
In the $\mbox{SU}(1,1)$ formalism the scalar product is :
\begin{eqnarray} 
q^{\alpha} \, p_{\alpha} \, & = & \, q_{\alpha}^{*} \, \eta^{\alpha \, \beta} \, p_{\beta} =  q^{*}_{1} p_{1} - 
q^{*}_{2} p_{2} \nonumber \\
q^{\alpha} \, q_{\alpha} \, & = & \, q_{\alpha}^{*} \, \eta^{\alpha \, \beta} \, q_{\beta} = |q_{1} |^2 -
|q_{2}|^2 \nonumber \\
A^{\alpha} \, q_{\alpha} \, & = & \,  A_{\alpha}^{*} \, \eta^{\alpha \, \beta} \, q_{\beta} 
\end{eqnarray}
we have introduced the vectors $q_{\alpha}$, $p_{\alpha}$, and the form $A^{\alpha}$ and so the vector $A_{\alpha}$ is :
\begin{equation}
A_{\alpha} \, = \, \eta_{\alpha \beta} A^{* \beta}
\end{equation}
\section{Notations and Conventions}
\label{notazie}
General adopted notations for first order actions are the
following ones:
\begin{eqnarray}
d & = & \mbox{ dimension of the world-volume $ \mathcal{W}_d$} \nonumber\\
D & = & \mbox{dimension of the bulk space--time $\mathcal{M}_D $ }\nonumber\\
V^{\underline{a}}&=& \mbox{vielbein $1$--form of bulk
space--time}\nonumber\\
\Pi^{\underline{a}}_{i} & = & D\times d \,\, \mbox{matrix. $0$--form auxiliary
field}\nonumber\\
h^{ij} &=& d\times d \,\,\mbox{symmetric matrix. $0$--form auxiliary
field}\nonumber\\
e^{\ell} &=& \mbox{vielbein $1$--form of the
world-volume}\nonumber\\
\eta_{\underline{ab}} & = & \mbox{diag}\{ +,\underbrace{-,\dots,-} _{D-1 \, times}
\} = \mbox{flat metric on the bulk}\nonumber\\
\eta^{\underline{ij}} & = & \mbox{diag} \{ +,\underbrace{-,\dots,-} _{d-1 \, times}
\} = \mbox{flat metric on the world--volume}\nonumber\\
\label{definizie}
\end{eqnarray}
The supersymmetric formulation of type IIB supergravity we rely on is
that of Castellani and Pesando \cite{igorleo} that uses the rheonomy
approach \cite{castdauriafre}. Hence, as it is customary in all the
rheonomy constructions, the adopted signature of space--time is the
\textbf{mostly minus signature}:
\begin{equation}
  \eta_{\underline{ab}}= \mbox{diag} \left\{ + ,\underbrace{
  -,\dots,-}_{\mbox{9 times}} \right \}
\label{mostmin}
\end{equation}
The index conventions are the following ones:
\begin{eqnarray}
\underline{a,b,c,\dots} & = & 0,1,2,\dots,9 \quad \mbox{Lorentz flat indices in $D=10$} \nonumber\\
i,j,k,\dots & = & 0,\dots ,d  \quad \mbox{Lorentz flat indices on the world-volume}\\
\alpha, \beta, \dots & = &  1,2  \quad \mbox{$\mathrm{SU(1,1)}$ doublet
indices}\nonumber\\
A,B,C,\dots & = & 1,2 \quad \mbox{$\mathrm{O(2)}$ indices for the scalar
coset}\nonumber\\
\Lambda,\Sigma,\Gamma,\dots &=& 1,2 \quad \mbox{$\mathrm{SL(2,R)}$ doublet
indices}
\label{indeconv}
\end{eqnarray}
For the gamma matrices our conventions are as follows:
\begin{equation}
  \left\{ \Gamma^{\underline{a}} \, , \, \Gamma^{\underline{b}} \right\}  = 2\,  \eta^{\underline{ab}}
\end{equation}
The convention for
constructing the dual of an $\ell$--form $\omega$ in D--dimensions is
the following:
\begin{equation}
  \omega= \omega_{\underline{i_1\dots i_\ell}} \, V^{\underline{i_1}} \wedge \dots \wedge
  V^{\underline{i_\ell}} \quad \Leftrightarrow \quad \star \omega = \frac{1}{(D-\ell)!} \,
  \epsilon_{\underline{a_1 \dots a_{D-\ell} b_1
  \dots b_\ell}} \omega^{\underline{b_1\dots
  b_\ell}} \, V^{\underline{a_1}} \wedge \dots \wedge V^{\underline{a_{D-\ell}}}
\label{hodgedual}
\end{equation}
Note that we also use $\ell$--form components with strength one:
$\omega= \omega_{\underline{i_1\dots i_\ell}} \, V^{\underline{i_1}} \wedge \dots \wedge
  V^{\underline{i_\ell}}$ and not with strength $\ell!$ as it would
  be the case if we were to write $\omega= \frac {1}{\ell !}\omega_{\underline{i_1\dots i_\ell}} \,
  V^{\underline{i_1}} \wedge \dots \wedge V^{\underline{i_\ell}}$
When it is more appropriate to use curved rather than flat indices
then the convention for Hodge duality is summarized by the formula:
\begin{equation}
  \star\left(dx^{\mu_1}\wedge\dots dx^{\mu_n}\right)=\frac{\sqrt{-{\rm det}(g)}}{(10-n)!}G^{\mu_1\nu_1}\dots
G^{\mu_n\nu_n}\epsilon_{\rho_1\dots
 \rho_{10-n}\nu_1\dots \nu_n}\,dx^{\rho_1}\wedge\dots dx^{\rho_{10-n}}
\label{hodgecurvo}
\end{equation}


\end{document}